\documentclass[pre,aps,twocolumn]{revtex4}

\usepackage{amssymb}
\usepackage{amsmath}

\usepackage{graphicx}
\usepackage{bm}                 

\usepackage[usenames]{color}
\usepackage{hyperref}
\usepackage[normalem]{ulem}

\usepackage{dcolumn}
\usepackage{array}

\usepackage{algorithm}
\usepackage{algpseudocode}
\usepackage{listings}
\usepackage{color} 

\definecolor{mygreen}{RGB}{28,172,0} 

\definecolor{mylilas}{RGB}{170,55,241}

\begin{document}

\lstset{language=Matlab,%
    breaklines=true,%
    morekeywords={matlab2tikz},
    keywordstyle=\color{blue},%
    morekeywords=[2]{1}, keywordstyle=[2]{\color{black}},
    identifierstyle=\color{black},%
    stringstyle=\color{mylilas},
    commentstyle=\color{mygreen},%
    showstringspaces=false,
    numbers=left,%
    numberstyle={\tiny \color{black}},
    numbersep=9pt, 
    emph=[1]{for,end,break},emphstyle=[1]\color{blue}, 
    emph=[2]{size,pinv,norm,max,abs,repmat},emphstyle=[1]\color{blue}, 
}

\title{Compressed Sensing by Shortest-Solution Guided Decimation}

\author{
  Mutian Shen$^{3}$, Pan Zhang$^{1}$, and Hai-Jun Zhou$^{1,2,4}$
}

\affiliation{
  $^1$Key Laboratory for Theoretical Physics, Institute of Theoretical Physics, Chinese Academy of Sciences, Beijing 100190, China \\
  $^2$School of Physical Sciences, University of Chinese Academy of Sciences, Beijing 100049, China \\
  $^3$School of the Gifted Young, University of Science and Technology of China, Hefei 230026, China \\
  $^4$Synergetic Innovation Center for Quantum Effects and Applications, Hunan Normal University, Changsha, Hunan 410081, China
}

\date{25 September, 2017 (version 1); 05 October, 2017 (version 2); 18 December, 2017 (version 3)}


\begin{abstract}
Compressed sensing is an important problem in many fields of science and engineering. It reconstructs signals by finding sparse solutions to underdetermined linear equations. In this work we propose a deterministic and non-parametric algorithm SSD (Shortest-Solution guided Decimation) to construct support of the sparse solution under the guidance of the dense least-squares solution of the recursively decimated linear equation. The most significant feature of SSD is its insensitivity to correlations in the sampling matrix. Using extensive numerical experiments we show that SSD greatly outperforms $\ell_1$-norm based methods, Orthogonal Least Squares, Orthogonal Matching Pursuit, and Approximate Message Passing when the sampling matrix contains strong correlations. This nice property of correlation tolerance makes SSD a versatile and robust tool for different types of real-world signal acquisition tasks.
\end{abstract}

\maketitle

\section{Introduction}

Real-world signals such as images, voice streams and text documents are highly compressible due to their intrinsic sparsity. Recent intensive efforts from diverse fields (computer science, engineering, mathematics and physics) have established the feasibility of merging data compression with data acquisition to achieve high efficiency of sparse information retrieval~\cite{Shi-etal-2009,Foucart-Rauhut-2013}. This integrated signal processing framework is called compressed sensing or compressed sampling~\cite{Candes-etal-2006,Donoho-2006,Gilbert-etal-2002}. At the core of this sampling concept is an underdetermined linear equation involving an  $M\!\times\!N$ real-valued matrix $\bm{D}$
\begin{equation}
  \bm{D} \bm{h} = \bm{z} \; .
  \label{eq:sparse02}
\end{equation}
(Throughout this paper we use uppercase and lowercase bold letters to denote matrices and vectors, respectively.) The column vector $\bm{h} \equiv (h_1, h_2, \ldots, h_N)^T$ is to be determined, while the column vector $\bm{z}\equiv (z_1, z_2, \ldots, z_M)^T$ is the result of $M$ sampling operations (measurements) on a hidden signal $\bm{h}^0\equiv (h_1^0, h_2^0, \ldots, h_N^0)^T$; in matrix form, this is
\begin{equation}
  \bm{z} \equiv \bm{D} \bm{h}^0 \; .
  \label{eq:measure}
\end{equation}
The vector $\bm{h}^0$ is called a \textit{planted solution} of (\ref{eq:sparse02}). Given a matrix $\bm{D}$ and an observed vector $\bm{z}$, the task is to reconstruct the planted solution $\bm{h}^0$.

The sampling process (\ref{eq:measure}) has compression ratio $\alpha\!\equiv\!\frac{M}{N}\!<\!1$. If the number $N_{nz}$ of non-zero entries in $\bm{h}^0$ exceeds $M$, some information must be lost in compression and it is then impossible to completely recover $\bm{h}^0$ from $\bm{z}$. However, if the sparsity $\rho \equiv \frac{N_{nz}}{N}$ is sufficiently below the compression ratio ($\rho < \alpha$),  $\bm{h}^0$ can be faithfully recovered by treating (\ref{eq:sparse02}) as the sparse representation problem of obtaining a solution $\bm{h}$ with the least number of non-zero entries~\cite{Donoho-Elad-2003, Zhang-etal-2015}.

As $\ell_0$-norm minimization is intractable, many different heuristic ideas have instead been explored for this sparse recovery task~\cite{Shi-etal-2009,Zhang-etal-2015}. These empirical methods form three major clusters: greedy deterministic algorithms for $\ell_0$-norm minimization, convex relaxations, and physics-inspired message-passing methods for approximate Bayesian inference. Representative and most popular algorithms of these categories are Orthogonal Least Squares (OLS) and Orthogonal Matching Pursuit (OMP)~\cite{Chen-etal-1989,Mallat-Zhang-1993,Pati-etal-1993,Davis-etal-1994,Tropp-2004,Tropp-Gilbert-2007,Dai-Milenkovic-2009,Chatterjee-etal-2012,Wang-Li-2017,Wen-etal-2017}, $\ell_1$-norm based methods~\cite{Candes-etal-2006,Donoho-Elad-2003,Gribonval-Nielsen-2003,Donoho-2006b,Tibshirani-1996,Chen-etal-2001}, and Approximate Message Passing (AMP)~\cite{Donoho-Maleki-Montanari-2009,Krzakala-etal-2011}, respectively. (In statistical physics the AMP method is known as the Thouless-Anderson-Palmer equation~\cite{Thouless-Anderson-Palmer-1977}.) To achieve good reconstruction performance, these algorithms generally assume the sampling matrix $\bm{D}$ to have the restricted isometric property (RIP)~\cite{Candes-Tao-2006}. This requires $\bm{D}$ to be sufficiently random,  uncorrelated and incoherent, which is not necessarily easy to meet in real-world applications. In many situations the matrix $\bm{D}$ is intensionally designed to be highly structured. For instance, $\bm{D}$ in the closely related dictionary learning problem often contains several complete sets of orthogonal base vectors to increase the representation capacity, leading to considerable correlations among the columns~\cite{Donoho-Elad-2003,Gribonval-Nielsen-2003}. An efficient and robust algorithm applicable for correlated sampling matrices is highly desirable.

In this work we introduce a simple deterministic algorithm, Shortest-Solution guided Decimation (SSD), that is rather  insensitive to the statistical property of the sampling matrix $\bm{D}$.  This algorithm has the same structure as the celebrated OLS and OMP algorithms~\cite{Chen-etal-1989,Mallat-Zhang-1993,Pati-etal-1993,Davis-etal-1994,Tropp-2004,Tropp-Gilbert-2007}. It selects a single column of $\bm{D}$ at each iteration step. The most significant difference is on how to choose the matrix columns: While OLS and OMP select the column that having the largest magnitude of (rescaled) inner product with the residual of vector $\bm{z}$, SSD selects a column under the guidance of the dense least-squares (i.e., shortest Euclidean-length) solution of the decimated linear equation. SSD exploits the left- and right-singular vectors of the decimated sampling matrix.

On random uncorrelated matrices SSD works better than OLS, OMP and $\ell_1$-norm based methods; it is slightly worse than AMP which for uncorrelated Gaussian matrices is proven to be Bayes-optimal under correct priors~\cite{Donoho-Maleki-Montanari-2009,Bayati-Montanari-2011}. A major advantage of SSD is that it does not require the sampling matrix to satisfy the RIP condition. Indeed we observe that the performance of SSD on highly correlated matrices remains the same even when $\ell_1$-norm based methods, OLS and OMP, and AMP all fail completely. This remarkable feature makes SSD a versatile and robust tool for different types of practical compressed sensing and sparse approximation tasks.

\section{The general dense solution}

When the number $M$ of constraints is less than the number of degrees of freedom $N$ ($\alpha\!<\!1$), the linear equation (\ref{eq:sparse02}) has infinitely many solutions and most of them are dense (all the entries of $\bm{h}$ are non-zero). To gain a geometric understanding on the general (dense) solution, we first express the sampling matrix in singular value decomposition (SVD)~\cite{Golub-Reinsch-1970} as $\bm{D} = \bm{U} \bm{\Lambda} \bm{V}^T$. Here $\bm{U}$ is an $M\!\times\!M$ orthonormal matrix; $\bm{\Lambda}$ is an $M\!\times\!N$ diagonal matrix whose diagonal elements are the $M$ non-negative singular values $\lambda_1 \geq \lambda_2 \geq \ldots \geq \lambda_M$; and $\bm{V}$ is an $N\!\times\!N$ orthonormal matrix. The columns of  $\bm{U} \equiv (\bm{u}^{1}, \bm{u}^{2}, \ldots, \bm{u}^{M})$ form a complete set of unit base vectors for the $M$-dimensional real space, so the vectors $\bm{u}^\mu \equiv (u_1^\mu, u_2^\mu, \ldots, u_M^\mu)^T$ satisfy $ \langle \bm{u}^{\mu}, \bm{u}^\nu \rangle\!=\!\delta_{\mu}^\nu$. Here and in following discussions $\langle \bm{a}, \bm{b}\rangle$ denotes the inner product of two vectors $\bm{a}$ and $\bm{b}$ (e.g., $\langle \bm{u}^\mu, \bm{u}^\nu \rangle\!\equiv\!\sum_{i} u_i^\mu u_i^\nu$); and $\delta_{\mu}^{\nu}$ is the Kronecker symbol, $\delta_{\mu}^{\nu}\!=\!0$ for $\mu\!\neq\!\nu$ and $\delta_{\mu}^{\nu}\!=\!1$ for $\mu\!=\!\nu$. Similarly, the orthonormal matrix $\bm{V} \equiv (\bm{v}^{1}, \bm{v}^{2}, \ldots, \bm{v}^{N})$ is formed by a complete set of unit base vectors $\bm{v}^\mu \equiv (v_1^\mu, v_2^\mu, \ldots, v_N^\mu)^T$ for the $N$-dimensional real space, with $\langle \bm{v}^\mu, \bm{v}^\nu \rangle = \delta_\mu^\nu$.

\begin{figure}
  \centering
  \includegraphics[width=0.6\columnwidth]{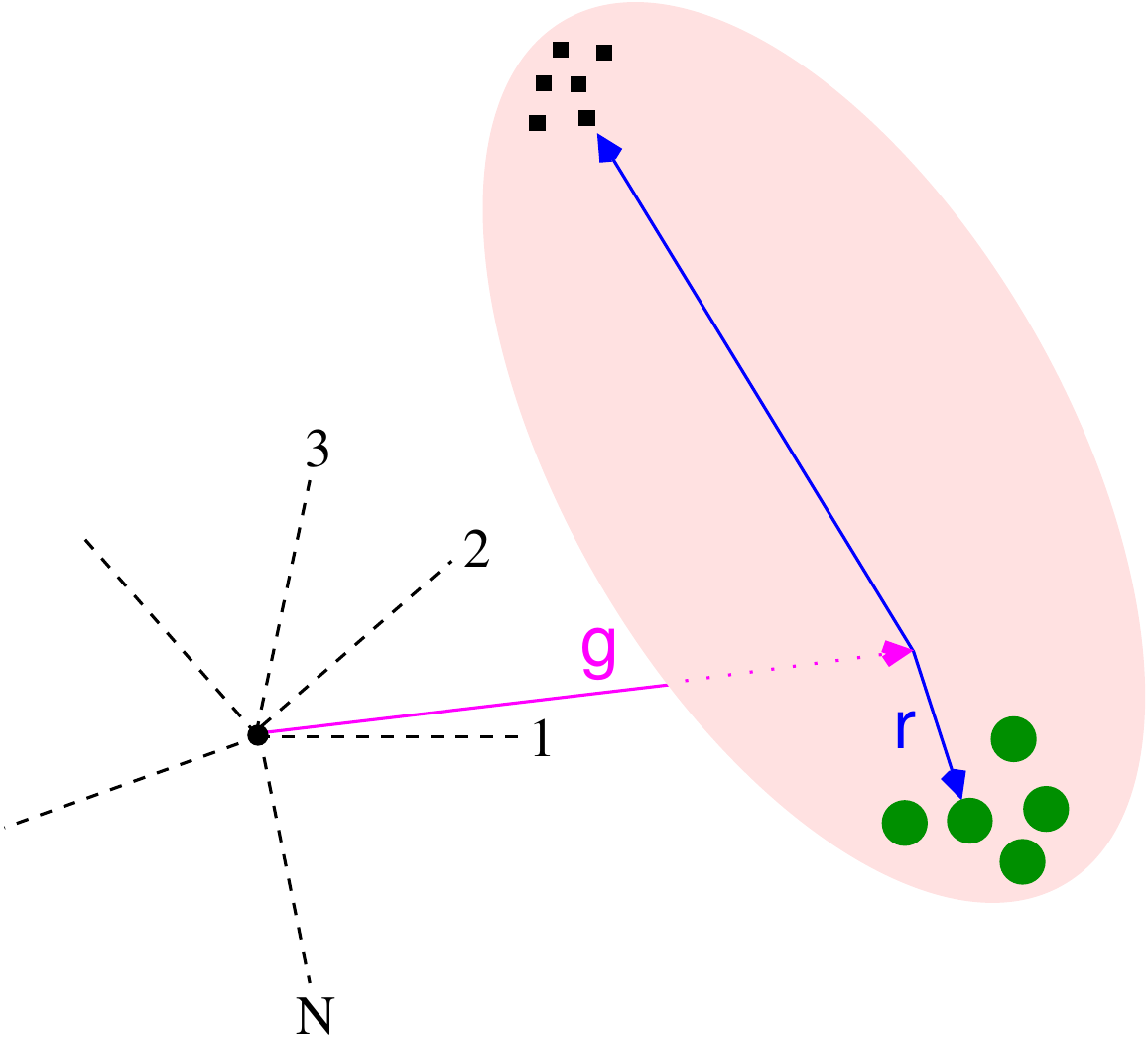}
  \caption{
    \label{fig:solutiong}
    Geometric interpretation on the general solution $\bm{h}$ in (\ref{eq:solutiong}). The guidance vector $\bm{g}\equiv (g_1, \ldots, g_N)^T$ lies within an $M$-dimensional subspace; the remainder vector $\bm{r}$ lies within the $(N\!-\!M)$-dimensional null space of matrix $\bm{D}$ signified by the elliptic shaded area, which is perpendicular to $\bm{g}$. Each dashed line represents one of the $N$ coordinate axes. In this schematic figure $g_1$ has the largest magnitude among all the entries of $\bm{g}$, so the leading index $l$ is $l=1$. The sparsest solution of (\ref{eq:sparse02}) is very likely to have large projection on the leading coordinate axis $l$ but is unlikely to have zero projection on this axis $l$. Filled circular and square points are examples of likely and unlikely sparse solutions, respectively.
  } 
\end{figure}

Expanding the vector $\bm{z}$ as a linear combination of base vectors $\bm{u}^\mu$, we obtain the following expression for the general solution of (\ref{eq:sparse02}):
\begin{equation}
  \bm{h} = \bm{g}  + \sum\limits_{\nu=M+1}^{N} c_\nu \bm{v}^{\nu} \; . 
  \label{eq:solutiong}
\end{equation}
Each of the $N\!-\!M$ coefficients $c_\nu$ is a free parameter that can take any real value; $\bm{g}\equiv (g_1, g_2, \ldots, g_N)^T$ is a column vector uniquely determined by $\bm{z}$:
\begin{equation}
  \bm{g} = 
  \sum\limits_{\mu=1}^M \Theta(\lambda_\mu)
  \frac{\langle \bm{z}, \bm{u}^{\mu}\rangle}{\lambda_\mu}
  \bm{v}^\mu
  \equiv  \bm{D}^{+} \bm{z}
  \; ,
 \label{eq:guidance}
\end{equation}
where $\Theta(x)$ is the Heaviside function, $\Theta(x)\!=\!1$ for $x\!>\!0$ and $\Theta(x)\!=\!0$ for $x\!\leq\!0$; $\bm{D}^+$ is the (Moore-Penrose) pseudo-inverse of $\bm{D}$~\cite{Golub-Reinsch-1970}. We call $\bm{g}$ the guidance vector.  We will rank the $N$ entries of $\bm{g}$ in descending order of their magnitude, and the index $l \in \{1, 2, \ldots, N\}$ of the maximum-magnitude entry $g_l$ of $\bm{g}$ will be referred to as the leading index.

The guidance vector $\bm{g}$ is perpendicular to all the base vectors $\bm{v}^{\nu}$ with  $\nu\!>\!M$. The remainder term $\bm{r}\!\equiv\!\sum_{\nu > M} c_\nu \bm{v}^\nu$ of the general solution (\ref{eq:solutiong}) therefore is perpendicular to $\bm{g}$.  A simple geometric interpretation of (\ref{eq:solutiong}) is illustrated in Fig.~\ref{fig:solutiong}: the solution $\bm{h}$ can move freely within the $(N\!-\!M)$-dimensional subspace (called the null space or kernel of $\bm{D}$) spanned by the base vectors $\bm{v}^{\nu}$ with $\nu\!>\!M$; this subspace is perpendicular to the guidance vector $\bm{g}$, which itself is a vector within the $M$-dimensional subspace spanned by the base vectors $\bm{v}^{\mu}$ with $\mu \leq M$.

Notice that the guidance vector $\bm{g}$ is nothing but the shortest Euclidean-length (i.e., least squares) solution~\cite{Golub-Reinsch-1970}. That is, the $\ell_2$-norm $||\bm{g}||_2 \equiv \sqrt{\sum_j g_j^2}$ achieves the minimum value among all the solutions of (\ref{eq:sparse02}). This is easily proved by setting all the coefficients $c_\nu$ of (\ref{eq:solutiong}) to zero. We may therefore compute $\bm{g}$ by the LQ decomposition method which is more economic than SVD. The sampling matrix is decomposed as $\bm{D} = \bm{L}\bm{Q}^T$, with $\bm{Q} \equiv (\bm{q}^1, \bm{q}^2, \ldots, \bm{q}^M)$ being an $N \times M$ orthonormal matrix (the columns $\bm{q}^\mu$ satisfying $\langle \bm{q}^\mu, \bm{q}^\nu\rangle = \delta_\mu^\nu$ for $\mu, \nu \in \{1,\ldots, M\}$) and $\bm{L}$ being an $M \times M$ lower-triangular matrix. The guidance vector $\bm{g}$ is then a linear combination of the unit vectors $\bm{q}^\mu$:
\begin{equation}
  \bm{g} = \sum\limits_{\mu=1}^{M} a_\mu \bm{q}^\mu \; , 
  \label{eq:guidanceLQ}
\end{equation}
and the coefficient vector $\bm{a} \equiv (a_1, \ldots, a_M)^T$ is easily fixed by solving the lower-triangular linear equation $\bm{L} \bm{a} = \bm{z}$.

The least-squares solution $\bm{g}$ can also be obtained through convex minimization with the following cost function
\begin{equation}
  \label{eq:convexmin}
  \frac{1}{2} \bm{g}^2 + \bm{\beta}^T  \bigl(\bm{z}-\bm{D} \bm{g}\bigr) \; ,
\end{equation}
where $\bm{\beta}\!\equiv\!(\beta_1, \beta_2, \ldots, \beta_M)^T$ is a column vector of $M$ Lagrange multipliers $\beta_i$~\cite{Boyd-etal-2011}. We can employ a simple iterative method of dual ascent:
\begin{subequations}
  \label{eq:dam}
  \begin{align}
    \bm{g}^{(t)} &= \bm{D}^T \bm{\beta}^{(t)} \; , \\
    \bm{\beta}^{(t+1)} &= \bm{\beta}^{(t)} + \varepsilon^{(t)}
    \bigl(\bm{z} - \bm{D} \bm{g}^{(t)}\bigr) \; .
  \end{align}
\end{subequations}
At each iteration step $t=0, 1, \ldots$ the updating rate parameter $\varepsilon^{(t)}$ is set to an optimal value to minimize the convergence time, see Appendix~\ref{sec:epsilon} for the explicit formula.

\section{The guidance vector as a cue}

The guidance vector $\bm{g}$ is dense and it is not the planted solution $\bm{h}^0$ we are aiming to reconstruct. But, does the dense $\bm{g}$ bring some reliable clues about the sparse $\bm{h}^0$? Because the observed vector $\bm{z}$ encodes the information of $\bm{h}^0$ through (\ref{eq:measure}), we see that
\begin{equation}
  \bm{g}  = 
  \sum\limits_{\mu=1}^{M} \langle \bm{h}^0, \bm{v}^\mu \rangle \bm{v}^\mu
  \equiv  \bm{P} \bm{h}^0 \; .
  \label{eq:h0vsg} 
\end{equation}
Here the $N\times N$ matrix $\bm{P}$ is a projection operator:
\begin{equation}
  \bm{P} \equiv \sum\limits_{\mu=1}^{M} 
  (v_1^\mu, v_2^\mu, \ldots, v_N^\mu)^T (v_1^\mu, v_2^\mu, \ldots, v_N^\mu) \; ,
  \label{eq:projectionP}
\end{equation}
which projects $\bm{h}^0$ to the subspace spanned by the first $M$ base vectors $\bm{v}^\mu$ ($1\!\leq\!\mu\!\leq\!M$) of matrix $\bm{V}$. A diagonal entry of $\bm{P}$ is $ P_{i i}  =  \sum_{\mu=1}^{M} (v_i^\mu)^2$, while the expression for an off-diagonal entry is $P_{i j}  =  \sum_{\mu=1}^{M} v_i^\mu v_j^\mu$.

Notice that, had the summation in (\ref{eq:projectionP}) included all the $N$ base vectors, $\bm{P}$ would have been the identity matrix ($P_{i j}\!=\!\delta_{i}^{j}$), and then $\bm{g}$ would have been identical to $\bm{h}^0$. Because only $M$ base vectors are included in (\ref{eq:projectionP}) the non-diagonal entries of $\bm{P}$ no longer vanish, but their magnitudes are still markedly smaller than those of the diagonal entries. Since $\bm{v}^\mu$ is a unit vector we may expect that $v_i^\mu\!\approx\!\pm\frac{1}{\sqrt{N}}$, then we estimate $P_{i i}\!\approx\!\frac{M}{N}\!=\!\alpha$. We may also expect that $v_i^\mu$ and $v_j^\mu$ are largely independent of each other, then we get that $P_{i j}\!\approx\!\pm\frac{\sqrt{M}}{N}$ (with roughly equal probability to be positive or negative).

Let us now focus on one entry $g_i$ of $\bm{g}$. According to (\ref{eq:h0vsg})
\begin{equation}
  g_i =   P_{i i} h_i^0 + \sum\limits_{n \neq i} P_{i n} h_n^0
  \equiv g_i^{(a)} + g_i^{(b)}  \; .
  \label{eq:gidecompose}
\end{equation}
Because $\bm{h}^0$ is a sparse vector with only $\rho N$ non-zero entries, the summation  in the above expression ($g_i^{(b)}$) contains only $\rho N$ terms. Neglecting the possible weak correlations among the coefficients $P_{i n}$, we get $g_i^{(b)}\!\approx\!\pm\frac{\sqrt{M}}{N}\sqrt{\rho N}a_0\!=\!\pm \sqrt{\alpha \rho}a_0$, where $a_0\!\equiv\!\sqrt{\frac{1}{\rho N}\sum_{i=1}^{N}(h_i^0)^2}$ is the rooted square mean value of the non-zero entries of $\bm{h}^0$. Notice that $g_i^{(b)}$ does not depend much on the index $i$ and it is of the same order as the first term of (\ref{eq:gidecompose}), which is $g_i^{(a)}\!\approx\!\alpha h_i^0$.

For the leading index $l$ of $\bm{g}$, since $g_l$ must have the maximum magnitude among all the entries $g_i$, we expect that $g_l^{(a)}$ will have the same sign as $g_l^{(b)}$ and it will have considerably large magnitude. It then follows that $h_l^0$ is very likely to be non-zero and also that $|h_l^0| \gtrsim a_0$.

\begin{figure}
  \centering
  \includegraphics[angle=270,width=1.0\columnwidth]{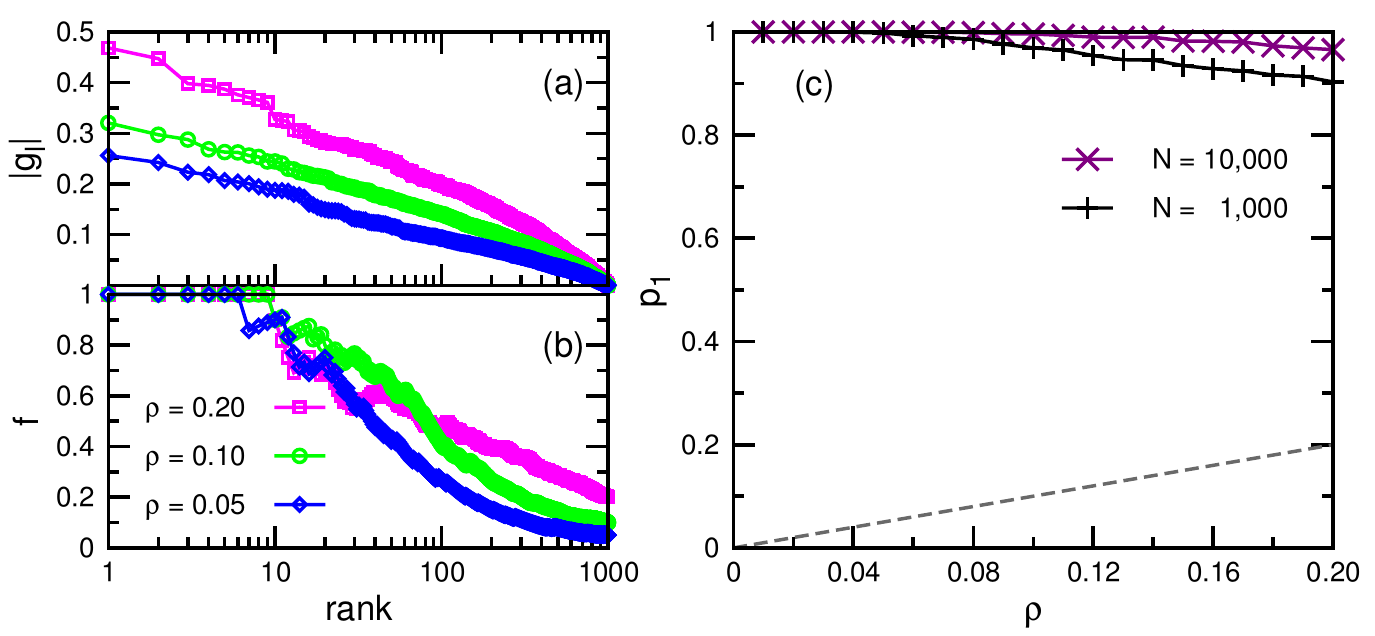}
  \caption{
    The guidance vector $\bm{g}$ reveals which entries of the planted solution $\bm{h}^0$ are most likely to be non-zero. Results are obtained on Gaussian matrices $\bm{D}$ with compression ratio $\alpha = 0.2$. (a): Rank curves for three guidance vectors of three $\bm{h}^0$ samples containing $\rho N$ non-zero entries with $\rho=0.05$, $0.1$ and $0.2$, respectively ($N=10^3$). The entries $g_i$ of $\bm{g}$ are ranked according to their magnitudes; the corresponding curves in (b) show the fractions $f(n)$ of non-zero entries of $\bm{h}^0$ among the $n$ top-ranked indexes $i$. (c): Probability $p_1$ of $h_l^0$ being non-zero ($l$ being the leading index of $\bm{g}$). $p_1$ is evaluated over $10^4$ random $\bm{h}^0$ samples, each of which containing $\rho N$ non-zero entries uniformly distributed in $(-1, +1)$, with$N=10^3$ (plus symbols) or $N=10^4$ (cross symbols). As a comparison, the dashed line shows the probability of an entry of $\bm{h}^0$ being non-zero if it is chosen uniformly at random among all the $N$ entries.
  }
  \label{fig:cues}
\end{figure}

The above theoretical analysis suggests that the vector $\bm{g}$ is very helpful for us to guess which entries of the planted solution $\bm{h}^0$ have large magnitudes. The validity of this theoretical insight has been confirmed by our numerical simulation results (Fig.~\ref{fig:cues}). We find that indeed the guidance vector $\bm{g}$ contains valuable clues about the non-zero entries of $\bm{h}^0$. If an index $i$ is ranked on the top with respect to the magnitude of $g_i$,  the corresponding value $h_i^0$ has a high probability to be non-zero (Fig.~\ref{fig:cues}a and \ref{fig:cues}b). This is especially true for the leading index $l$ of $\bm{g}$. We also observe that, both for $\rho\!<\!\alpha$ and for $\rho\!\simeq\!\alpha$, the probability of $h_l^0$ being non-zero becomes more and more close to unity as the size $N$ increases (Fig.~\ref{fig:cues}c). This indicates the cue offered by $\bm{g}$ is more reliable for larger-sized compressed sensing problems.

Building on the theoretical insights of (\ref{eq:h0vsg}) and (\ref{eq:gidecompose}), we now propose a simple algorithm for solving equation (\ref{eq:sparse02}).

\section{Shortest-Solution guided Decimation (SSD)}

Let us denote the sampling matrix as a collection of column vectors $\bm{d}^\mu$, i.e.,  $\bm{D}\!\equiv\!(\bm{d}^1, \bm{d}^2, \ldots, \bm{d}^N)$. With respect to the leading index $l$ of $\bm{g}$ the linear equation (\ref{eq:sparse02}) is rewritten as
\begin{equation}
  h_{l} \bm{d}^l + \sum\limits_{\mu \neq l} h_\mu \bm{d}^\mu  = \bm{z} \; .
  \label{eq:firststep}
\end{equation}
Therefore, if all the other $N\!-\!1$ entries $h_\mu$ of the vector $\bm{h}$ are known, $h_l$ is uniquely determined as
\begin{equation}
  h_{l} =
  \frac{\langle\bm{z}, \bm{d}^l\rangle}{\langle\bm{d}^{l}, \bm{d}^{l}\rangle}
  - \sum\limits_{\mu \neq l} h_\mu 
  \frac{\langle\bm{d}^{\mu}, \bm{d}^{l}\rangle}
       {\langle\bm{d}^{l}, \bm{d}^{l}\rangle} \; .
       \label{eq:xi0}
\end{equation}
Let us denote by $\bm{h}^{\backslash l}\!\equiv\!(h_1, \ldots, h_{l-1}, h_{l+1}, \ldots, h_{N})^T$ the vector formed by deleting $h_l$ from $\bm{h}$. This vector must satisfy the following linear equation
\begin{equation}
  \bm{B} \bm{h}^{\backslash l} = \bm{y} \; .
  \label{eq:Bxy}
\end{equation}
$\bm{B} \equiv (\bm{b}^1, \ldots, \bm{b}^{l-1}, \bm{b}^{l+1}, \ldots, \bm{b}^{N})$ is an $M\times (N\!-\!1)$ matrix decimated from $\bm{D}$ with its column vectors $\bm{b}^\mu$ being
\begin{equation}
  \bm{b}^\mu \equiv 
  \bm{d}^\mu - \frac{\langle \bm{d}^\mu, \bm{d}^{l}\rangle}
     {\langle \bm{d}^{l}, \bm{d}^{l}\rangle} \bm{d}^{l} 
     \quad  \quad\quad (\mu \neq l) \; .
     \label{eq:Bmtx}
\end{equation}
The $M$-dimensional vector $\bm{y}$ is the residual of $\bm{z}$:
\begin{equation}
  \bm{y} \equiv \bm{z} -
  \frac{\langle \bm{z}, \bm{d}^l\rangle}
       {\langle \bm{d}^l, \bm{d}^l\rangle} \bm{d}^l  \; .
       \label{eq:zvector}
\end{equation}
This residual vector $\bm{y}$ and all the decimated column vectors $\bm{b}^\mu$ are perpendicular to $\bm{d}^l$. 

Equation~(\ref{eq:Bxy}) has the identical form as the original linear problem (\ref{eq:sparse02}). If $\bm{y}$ is a zero vector, we can simply set $\bm{h}^{\backslash l}$ to be a zero vector too and then a solution $\bm{h}$ with a single non-zero entry $h_l$ is obtained by (\ref{eq:xi0}). On the other hand if the residual $\bm{y}$ is non-zero, we can obtain the shortest Euclidean-length (least squares) solution of (\ref{eq:Bxy}) as the new guidance vector. (If we adopt the dual ascent method (\ref{eq:dam}) for this task, it is desirable to start the iteration from the old guidance vector to save convergence time.) A new leading index (say $m$) will then be identified, and the corresponding entry $h_m$ is expressed by the remaining $N\!-\!2$ entries of $\bm{h}$ as
\begin{equation}
  h_m =
  \frac{\langle \bm{y}, \bm{b}^m\rangle}{\langle \bm{b}^m, \bm{b}^m\rangle}
  -\sum\limits_{\mu \neq l, m} h_\mu \frac{\langle \bm{b}^\mu, \bm{b}^m\rangle}{
    \langle \bm{b}^m, \bm{b}^m \rangle} \; .
  \label{eq:xm0}
\end{equation}

This Shortest-Solution guided Decimation (SSD) process will terminate within a number $K$ of steps ($K\!\leq\!M$). We will then achieve a unique solution for the $K$ selected entries $h_l, h_m, \ldots$ by backtracking the $K$ derived equations such as (\ref{eq:xm0}) and (\ref{eq:xi0}), setting all the other $N\!-\!K$ entries of $\bm{h}$ to be exactly zero. This solution $\bm{h}$ will have at most $K$ non-zero entries. We provide the pseudo-code of SSD in Algorithm ~\ref{alg:ssd} and the corresponding MATLAB code in Appendix~\ref{sec:appendix}. The longer {\tt C++} implementation is provided at {\tt power.itp.ac.cn/\~{}zhouhj/codes.html}.

\begin{algorithm}[t]
  \caption{
    \label{alg:ssd}
    Shortest-Solution guided Decimation (SSD) for the compressed sensing problem $\bm{D} \bm{h} = \bm{z}$.}
  \begin{algorithmic}[100]
    \State
    \textbf{Input}: $M\times N$ matrix $\bm{D} \equiv  (\bm{d}^1, \bm{d}^2, \ldots, \bm{d}^N)$; $M$-dimensional vector $\bm{z}$; index set $\Phi=\{1, 2, \ldots, N\}$; convergence threshold $\epsilon$ (set to be $10^{-8}$); stack $T$ of linear equations (initially $T=\emptyset$).
    \State
    \textbf{Output}: solution vector $\bm{h}=(h_1, h_2, \ldots, h_N)^T$.
    
    \State
    
    \While{$\frac{1}{M} \sum_{i=1}^{M} |z_i| > \epsilon$} \Comment{decimation}
    \begin{enumerate}
    \item[1.]
      Get the least-squares solution  $\bm{g} = \{g_\mu: \mu \in \Phi\}$ for the linear equation $\sum_{\mu \in \Phi} g_\mu \bm{d}^\mu = \bm{z}$.
    \item[2.]
      Get leading index $l$ of $\bm{g}$ by criterion $|g_l| \geq |g_\mu|$ ($\forall \mu\in \Phi$); then delete index $l$ from set $\Phi$.
    \item[3.] Deposit to stack $T$ the linear equation for $h_l$:
      $$
      h_l =    \frac{\langle\bm{z}, \bm{d}^l\rangle}{\langle\bm{d}^{l},
        \bm{d}^{l}\rangle}
      - \sum\limits_{\mu \in \Phi} h_\mu 
      \frac{\langle\bm{d}^{\mu}, \bm{d}^{l}\rangle}
           {\langle\bm{d}^{l}, \bm{d}^{l}\rangle} \; .
           $$
         \item[4.]
           Update $\bm{d}^\mu$ (for every $\mu \in \Phi$) and $\bm{z}$ as:
           \begin{align}
             \bm{d}^\mu & \leftarrow 
             \bm{d}^\mu - 
             \frac{\langle \bm{d}^\mu, \bm{d}^{l}\rangle}
                  {\langle \bm{d}^{l}, \bm{d}^{l}\rangle} \bm{d}^{l}\; , 
                  \quad \quad \mu \in \Phi \nonumber  \\
                  \bm{z} & \leftarrow 
                  \bm{z} - \frac{\langle \bm{z}, \bm{d}^l\rangle}
                     {\langle \bm{d}^l, 
                       \bm{d}^l\rangle} \bm{d}^l \; . \nonumber
           \end{align}
    \end{enumerate}
    \EndWhile
    \State Set $h_\mu = 0$ for every $\mu \in \Phi$.
    \While{Stack $T\neq \emptyset$} \Comment{backtrack}
    \begin{enumerate}
    \item[1.] Solve the top equation of stack $T$.
    \item[2.] Pop the top equation out of stack $T$.
    \end{enumerate}
    \EndWhile
   \end{algorithmic}
\end{algorithm}

\begin{figure}
\centering
\includegraphics[angle=270,width=1.0\columnwidth]{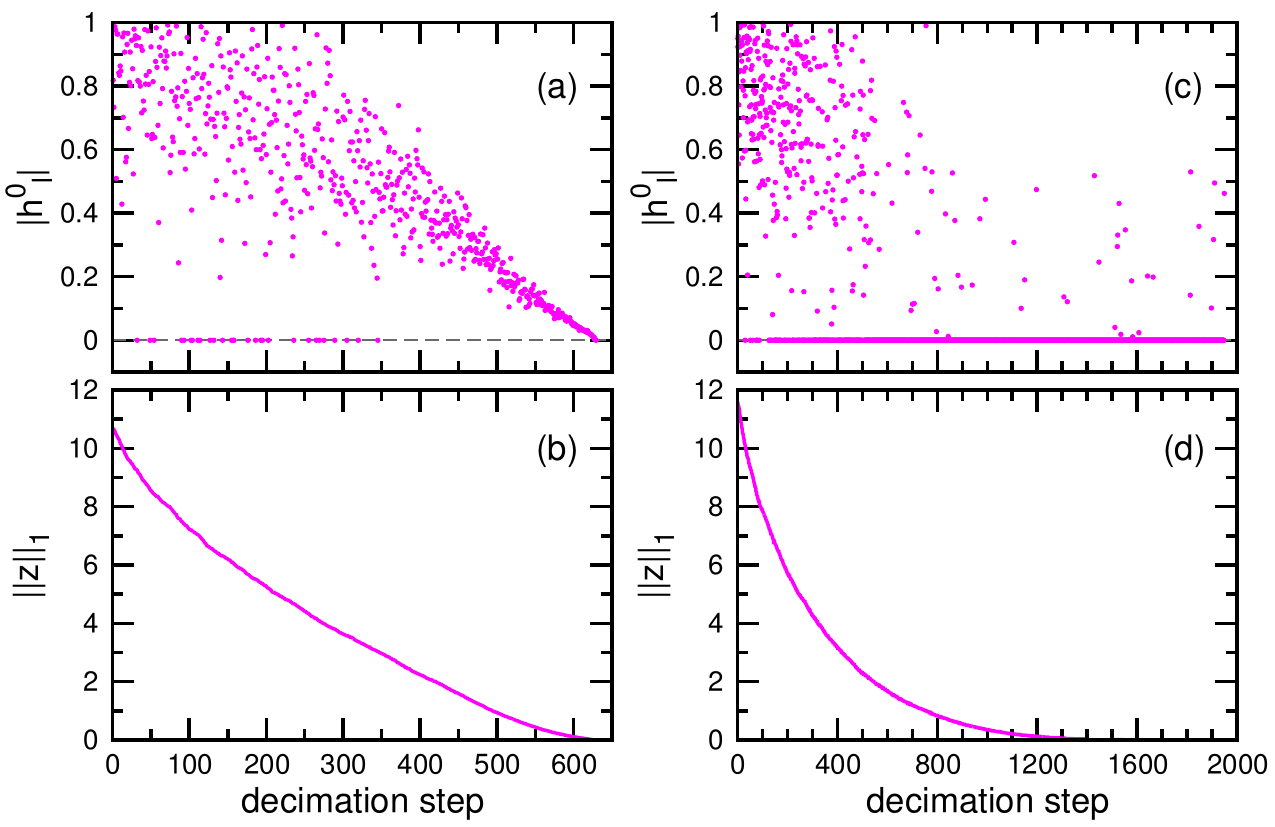}
\caption{
\label{fig:evolution}
Two traces of the SSD process on a single Gaussian matrix with $M=2000$ rows and $N=10000$ columns. The planted solutions $\bm{h}^0$ for the left [(a), (b)] and right [(c), (d)] columns are different, but they both have the same number $N_{nz}= 600$ of non-zero entries (each of them is an i.i.d random value uniformly distributed within $(-1, 1)$). The horizontal axis denotes the step of the decimation. $|h_l^0|$ is the $l$-th entry of $\bm{h}^0$ with $l$ being the identified leading index at this decimation step. $||{\bf z}||_1 = \sum_{i=1}^M |z_i|$ is the $\ell_1$-norm of the residual signal vector $\bm{z}$. For the left column, the decimation stops after $K=629$ steps with all the non-zero entries of $\bm{h}^0$ being identified (the $29$ mistakes are then all corrected by the followed backtrack process). For the right column, the decimation stops after $K=1948$ steps with only $374$ non-zero entries of $\bm{h}^0$ being identified (the followed backtrack process reports a solution that is dramatically different from $\bm{h}^0$).}
\end{figure}

An an illustration, we show in Fig.~\ref{fig:evolution} the traces of two SSD processes obtained on a $2000\!\times\!10000$ Gaussian matrix $\bm{D}$ for two planted solutions $\bm{h}^0$ with $600$ non-zero entries. We see that the leading index $l$ is not always reliable, sometimes the true value of $h_l^0$ is actually zero. But the SSD algorithm is robust to these guessing mistakes. As long as the number of such mistakes is relatively small they will all be corrected by the backtrack process of Algorithm~\ref{alg:ssd} (Fig.~\ref{fig:evolution}a and \ref{fig:evolution}b). But if these guessing mistakes are too numerous   (Fig.~\ref{fig:evolution}c and \ref{fig:evolution}d), the backtrack process is unable to correct all of them and the resulting solution $\bm{h}$ will be dense. For Algorithm~\ref{alg:ssd} to be successful, the only requirement is that the indexes of all the non-zero entries of $\bm{h}^0$ are removed from the working index set $\Phi$ in no more than $M$ steps. If this condition is satisfied the backtrack process of the algorithm will then completely recover the planted solution $\bm{h}^0$.

The SSD algorithm has the same structure as the greedy OLS~\cite{Chen-etal-1989} and OMP~\cite{Pati-etal-1993,Davis-etal-1994} algorithms. By carefully comparing these three algorithms we realize that they differ only in how to construct the guidance vector $\bm{g}$. At the beginning of the $n$-th decimation step ($n=0, 1, \ldots$) the updated sampling matrix $\bm{D}$ has dimension $M\!\times\!(N-n)$. Then OMP simply sets $\bm{g}\!=\!\bm{D}^T \bm{z}$ so the $\mu$-th entry is $g_\mu\!=\!\langle \bm{z}, \bm{d}^\mu\rangle$, where $\bm{d}^\mu$ is the $\mu$-th column of the current matrix $\bm{D}$ and $\bm{z}$ is the current residual vector. In OLS, $g_\mu$ is computed by a slightly different expression $g_\mu\!=\!\frac{\langle \bm{z}, \bm{d}^\mu\rangle}{\sqrt{\langle \bm{d}^\mu, \bm{d}^\mu\rangle}}$. In SSD, $\bm{g}$ is the least-squares solution of $\bm{D} \bm{g}\!=\!\bm{z}$, so $\bm{g}\!=\!\bm{D}^+\bm{z}$ according to (\ref{eq:guidance}). After the leading index $l$ of $\bm{g}$ is determined, these algorithms then modify $\bm{D}$ and $\bm{z}$ by the same rule (see the two equations in step-$4$ of the \textit{while} loop of Algorithm~\ref{alg:ssd}). The total number of arithmetic operations performed in the $n$-th decimation step is approximately $6 M (N-n)$ for OMP and $8 M(M-n)$ for OLS; and for SSD this number is approximately $4 (L+1) M (N-n)$ if we use $L$ repeats of the iteration (\ref{eq:dam}) to update $\bm{g}$. It may be reasonable to assume $L\leq 10$, then SSD is roughly $5$--$7$ times slower than OMP and OLS. On the other hand, the results in the next section will demonstrate that the guidance vector of SSD is much better than those of OMP and OLS, especially on correlated sampling matrices.

\section{Comparative results}

We now test the performance of SSD on sparse reconstruction tasks involving both uncorrelated and correlated sampling matrices. As the measure of correlations we consider the condition number $Q \equiv \frac{\lambda^{max}}{\lambda^{min}}$, which is the ratio between the maximum ($\lambda^{max}$) and the minimum ($\lambda^{min}$) singular value of the sampling matrix $\bm{D}$. An uncorrelated matrix has condition number $Q \approx 1$, while a highly correlated or structured matrix has condition number $Q \gg 1$. Notice that the RIP condition is severely violated for matrices with large $Q$ values. Two types of sampling matrix are examined in our simulations:

\textbf{Gaussian:}
Each entry of the matrix is an identically and independently distributed (i.i.d) random real value drawn from the Gaussian distribution with mean zero and unit variance. Every two different columns $\bm{d}^\mu$ and $\bm{d}^\nu$ of such a random matrix are almost orthogonal to each other; that is, $\frac{1}{M} \bigl|\langle \bm{d}^\mu ,  \bm{d}^\nu\rangle\bigr| \approx \frac{1}{\sqrt{M}}$. The condition number $Q$ of a Gaussian matrix is close to $1$, and the RIP condition is satisfied.

\textbf{Correlated:}
The matrix $\bm{D}$ is obtained as the product of an $M\times R$ matrix $\bm{D}_1$ and a $R\times N$ matrix $\bm{D}_2$, so $\bm{D} = \bm{D}_1 \bm{D}_2$. Both $\bm{D}_1$ and $\bm{D}_2$ are Gaussian random matrices as described above. As the rank number $R$ approaches $M$ from above, the entries in the composite matrix $\bm{D}$ are more and more correlated and the condition number $Q$ increases quickly.

The planted solutions $\bm{h}^0$ are sparse random vectors with a fraction $\rho$ of non-zero entries. The positions of the $\rho N$ non-zero entries are chosen completely at random from the $N$ possible positions. Two types of planted solutions (Gaussian or uniform) are employed. The non-zero entries of $\bm{h}^0$ are i.i.d random real values drawn from the Gaussian distribution with mean zero and unit variance or from the uniform distribution over $(-1, +1)$, for the Gaussian-type and uniform-type $\bm{h}^0$, respectively. The relative distance $\Delta$ between $\bm{h}^0$ and the reconstructed solution $\bm{h}$ is defined as
\begin{equation}
 \Delta \equiv \frac{|| (\bm{h}-\bm{h}^0)||_2}{||\bm{h}^0||_2} \; .
\end{equation}
If this relative distance is considerably small (that is, $\Delta \leq 10^{-5}$), we claim that the planted solution $\bm{h}^0$ has been correctly recovered by the algorithm. We fix the compression ratio to $\alpha = 0.2$ in the numerical experiments. Two different values of $N$ are considered, $N=10^3$ and $N=10^4$.

Many heuristic algorithms have been designed to solve the compressed sensing problem. Here we choose four representative algorithms for the comparative study, namely Minimum $\ell_1$-norm (ML1), OLS, OMP, and AMP. The algorithm ML1 tries to find a solution of (\ref{eq:sparse02}) with minimized $\ell_1$-norm $\sum_{i=1}^N |h_i|$. Here we use the MATLAB toolbox L1\_MAGIC~\cite{candes2005l1}. The stop criterion for the primal-dual routine of this code is set to be $pdtol=10^{-3}$ following the literature.  The OLS and OMP algorithms are implemented according to \cite{Chen-etal-1989,Tropp-2004,Tropp-Gilbert-2007}. The AMP algorithm is implemented according to \cite{Donoho-Maleki-Montanari-2009,Krzakala-etal-2011}. The Gauss-Bernoulli prior is assumed for the planted solution $\bm{h}^0$ in the AMP algorithm as in \cite{Donoho-Maleki-Montanari-2009,Krzakala-etal-2011}, and the sparsity $\rho$ of $\bm{h}^0$ is revealed to AMP as input.

\subsection{Gaussian sampling matrix}

\begin{figure}
\centering
  \includegraphics[angle=270,width=1.0\columnwidth]{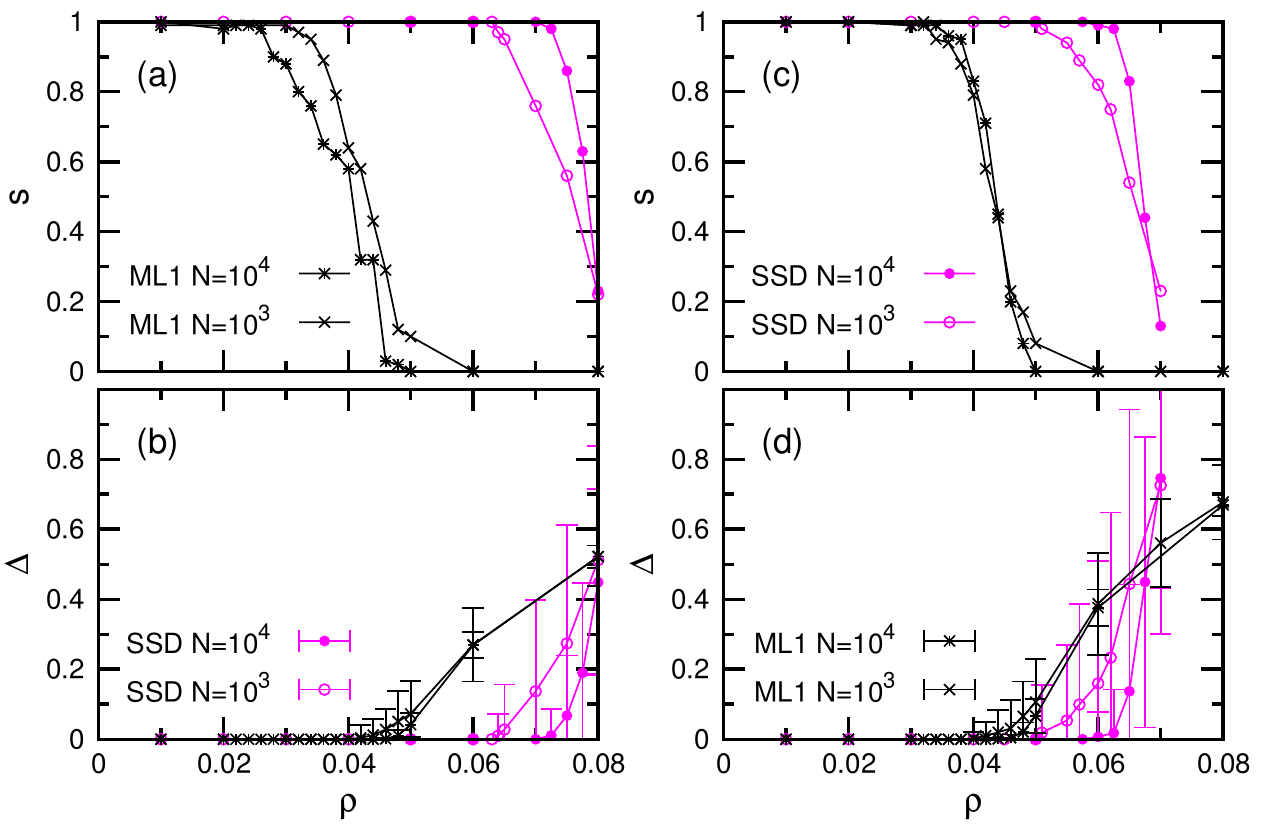}
  \caption{
  \label{fig:CompML1}
   Comparing the SSD and ML1 algorithms on two Gaussian sampling matrices $\bm{D}$ with compression level $\alpha=0.2$ ($N=10^3$ and $10^4$, respectively). Left column [(a), (b)] and right column [(c), (d)] correspond to Gaussian-type and uniform-type planted solutions $\bm{h}^0$, respectively. Each data point is obtained by averaging over $100$ samples of $\bm{h}^0$ with the same sparsity $\rho$. In (a) and (c),  $s$ is the fraction of successfully reconstructed solutions among the $100$ input $\bm{h}^0$ samples. Correspondingly, $\Delta$ in (b) and (d) is the relative distance (mean and standard deviation) between $\bm{h}^0$ and the reconstructed solution $\bm{h}$.
    }
\end{figure}

\begin{figure}
  \centering
  \includegraphics[angle=270,width=1.0\columnwidth]{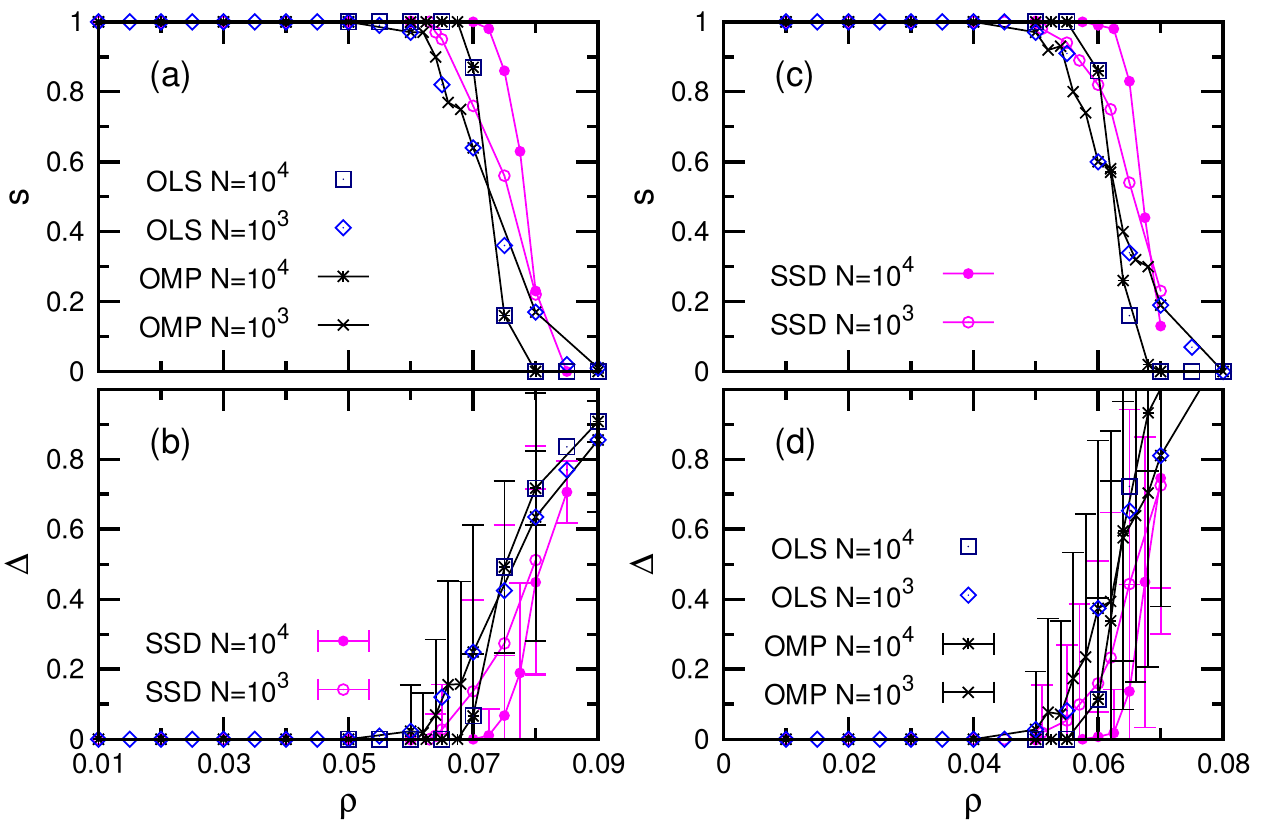}
  \caption{\label{fig:CompOMP}
    Comparing the performances of SSD with OLS and OMP on Gaussian random matrices. The matrices and the planted solutions are the same as those used in Fig.~\ref{fig:CompML1}. OLS has almost identical performance as OMP, so in (b) and (d) the standard deviations of the OLS results are not marked.
  }
\end{figure}

\begin{figure}
  \centering
  \includegraphics[angle=270,width=1.0\columnwidth]{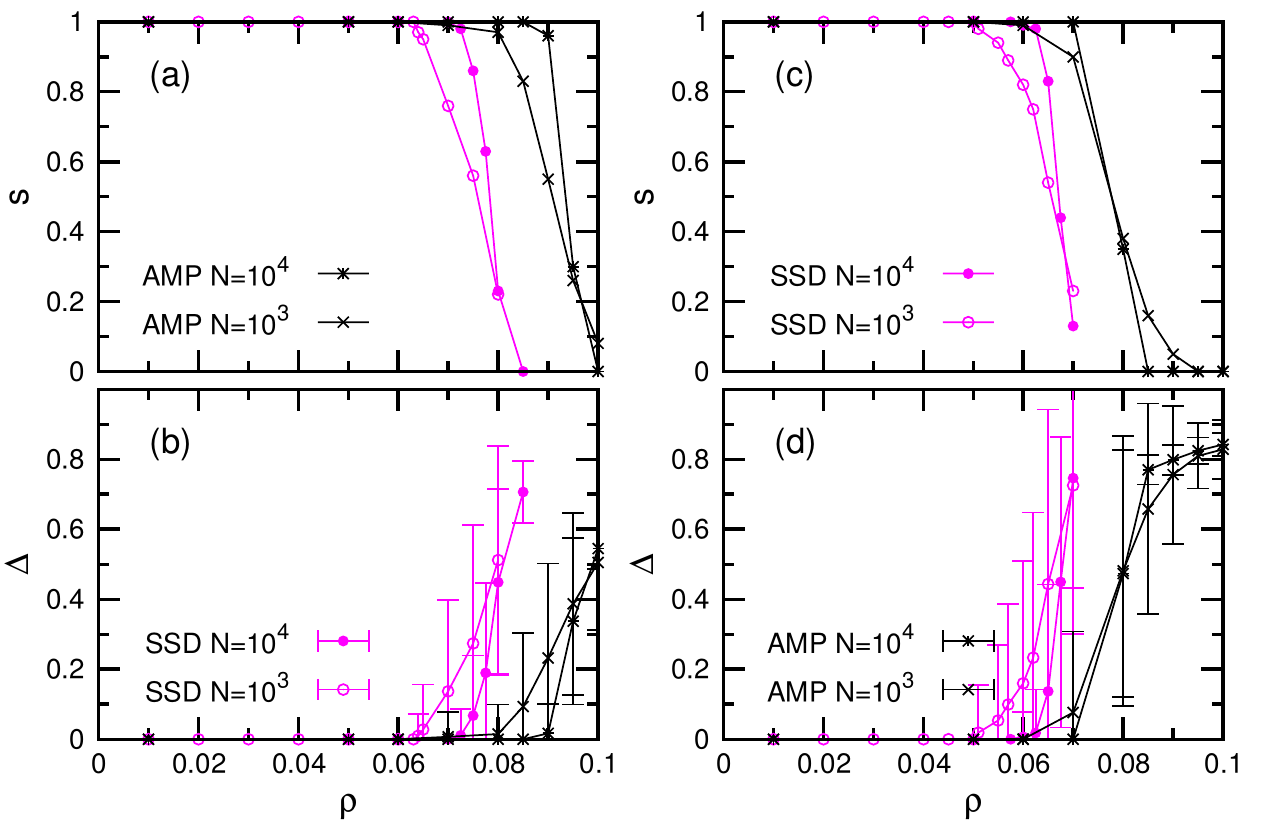}
  \caption{
    \label{fig:CompAMP}
    Comparing the performances of SSD and AMP on Gaussian random matrices. The matrices and the planted solutions are the same as those used in Fig.~\ref{fig:CompML1}.}
\end{figure}

The performance of SSD on Gaussian sampling matrices is compared with those of ML1 (Fig.~\ref{fig:CompML1}), OLS and OMP (Fig.~\ref{fig:CompOMP}), and AMP (Fig.~\ref{fig:CompAMP}).  Our simulation results demonstrate that, at fixed compression ratio $\alpha$  the transition between successful and unsuccessful reconstruction becomes sharper as $N$ increases. At $N=10^4$ and $\alpha=0.2$, when the planted solutions are Gaussian-type, SSD can successfully reconstruct the planted solutions with high probability (e.g., success rate $s> 50\%$) if the signal sparsity is $\rho \leq 0.078$. The corresponding values for ML1, OLS and OMP, and AMP are, respectively, $\rho \leq 0.04$, $0.07$ and $0.093$. When the planted solutions are uniform-type, SSD can successfully reconstruct at sparsity $\rho \leq 0.067$. The corresponding values for ML1, OLS and OMP, and AMP are, respectively, $\rho \leq 0.045$, $0.065$ and $0.079$.

It is encouraging to observe that SSD achieves much better performance than the $\ell_1$-norm based ML1. In comparison with ML1, the SSD algorithm is closer to $\ell_0$-norm minimization since it tries to find the non-zero entries of $\bm{h}^0$ but does not care about the magnitude of these entries.

The SSD algorithm slightly outperforms OLS and OMP on the Gaussian matrices. Because the different columns of a Gaussian random matrix $\bm{D}$ are almost orthogonal to each other, the pseudo-inverse $\bm{D}^+$ does not differ very much from the transpose $\bm{D}^T$. This could explain why SSD only weakly improves over OLS and OMP. Similar small improvements of SSD over OLS and OMP are observed on  uniform sampling matrices whose entries are i.i.d real values uniformly distributed in $(-1, +1)$.  These comparative results suggest that the SSD iteration process makes less guessing mistakes than the OLS and OMP iterations do.

The AMP algorithm performs considerably better than SSD on the Gaussian matrices (Fig.~\ref{fig:CompAMP}). This may not be surprising since AMP is an global optimization approach. Compared with SSD, the AMP algorithm also takes more additional information of $\bm{h}^0$ as input, including its sparsity level $\rho$ and the probability distribution of its non-zero entries, which may not be available in some real-world applications.

\subsection{Correlated sampling matrix}

\begin{figure}
  \centering
  \includegraphics[angle=270,width=0.7\columnwidth]{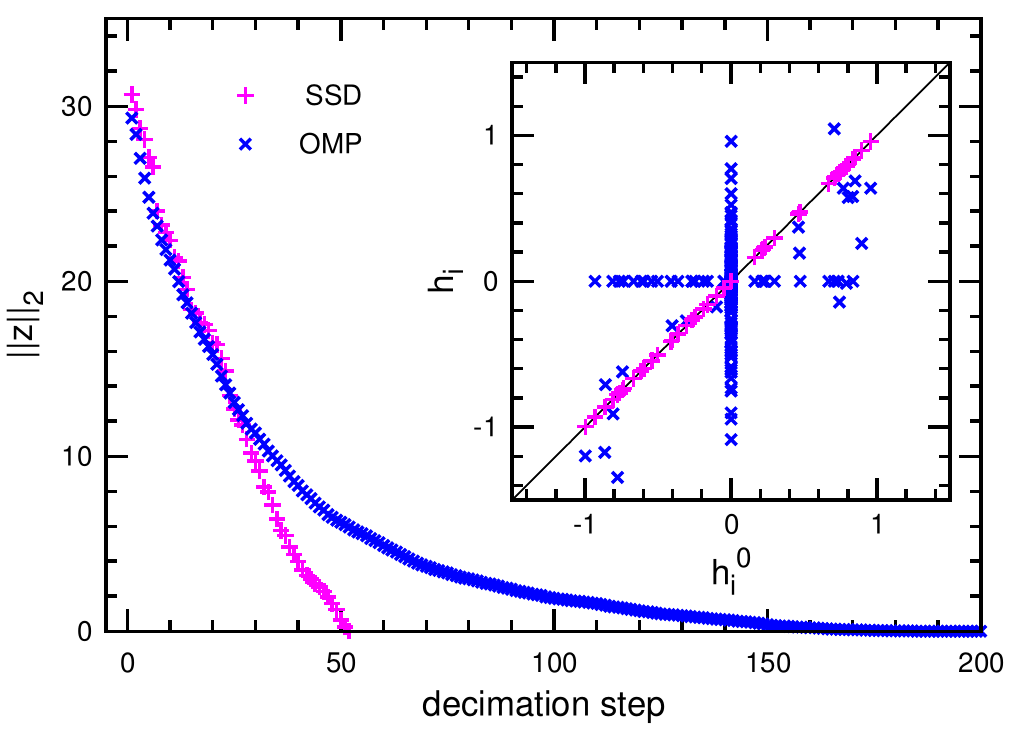}
  \caption{
    \label{fig:iterations}
    Comparing SSD and OMP on a single correlated measurement matrix $\bm{D}$ with $N=1000$ columns and $M=280$ rows. The protocol of generating this matrix  is described in the main text. The planted solution $\bm{h}^0$ is a uniform-type random sparse vector with $N_{nz}=50$ non-zero entries. The main figure shows how the $\ell_2$-norm $||\bm{z}||_2$ of the residual signal vector $\bm{z}$ changes with the decimation step (plus points, SSD results; cross points, OMP results). The inset compares the entries $h_i^0$ of the planted solution $\bm{h}^0$ and the corresponding entries $h_i$ of the recovered solution $\bm{h}$.
  }
\end{figure}

Although SSD in comparison with AMP does not give very impressive results on Gaussian random matrices, it works much better than AMP (and also OLS, OMP and ML1) on correlated matrices. Let us first consider an ill-conditioned $280\!\times\!1000$  matrix instance obtained by multiplying two random Gaussian matrices of size $280\!\times\!200$ and $200\!\times\!1000$. The condition number of this matrix is $Q=\infty$. The message-passing iteration of AMP quickly diverges on this instance and the algorithm then completely fails, even for very small sparsity $\rho=0.005$.  Similar divergence problem is experienced for the ML1 algorithm. OLS and OMP have no difficulty in constructing a solution $\bm{h}$, but their solutions are rather dense and are much different from the planted solution $\bm{h}^0$ (Fig.~\ref{fig:iterations}, cross points). On the other hand, SSD is not affected by the strong correlations of this matrix and it fully reconstructs $\bm{h}^0$ (Fig.~\ref{fig:iterations}, plus points). From Fig.~\ref{fig:iterations} we see that SSD initially causes a smaller decrease in the Euclidean length ($\ell_2$-norm) of the signal vector $\bm{z}$ than OMP (or OLS) does, but the slope of $||\bm{z}||_2$ decreasing keeps roughly the same in later iterations.

\begin{figure}
  \centering
  \includegraphics[angle=270,width=1.0\columnwidth]{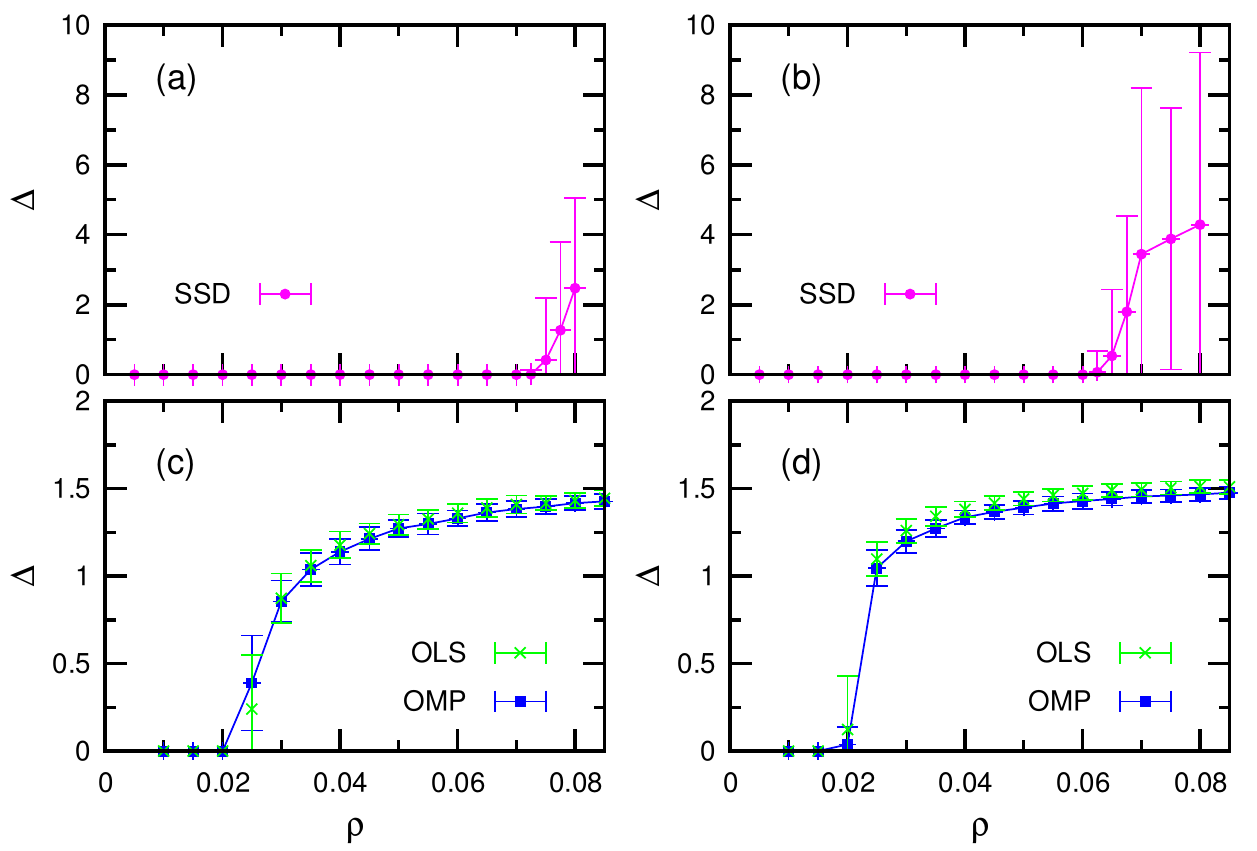}
  \caption{
    \label{fig:Cmtx10k}
    Comparing the performances of SSD [(a), (b)] and OLS and OMP [(c), (d)] on a highly correlated sampling matrix $\bm{D}$ with $N=10^4$ columns, compression level $\alpha=0.2$ and condition number $Q=3.84\times 10^5$.  Left column [(a), (c)] corresponds to Gaussian-type planted solutions $\bm{h}^0$, right column [(b), (d)] corresponds to uniform-type $\bm{h}^0$. $\rho$ is the sparsity of $\bm{h}^0$; $\Delta$ is the relative distance (mean and standard deviation, over $100$ random $\bm{h}^0$ samples) between $\bm{h}^0$ and the reconstructed solution $\bm{h}$.
  }
\end{figure}

Figure~\ref{fig:Cmtx10k} shows how the performances of SSD, OLS and OMP change with the sparsity $\rho$ of the planted solutions $\bm{h}^0$ on another larger matrix of $N=10^4$, $\alpha=0.2$, and condition number $Q=3.84\times 10^5$. (The results of AMP and ML1 are not shown because they always fail to recover $\bm{h}^0$.) We observe that OLS and OMP work only for sparsity $\rho < 0.02$, while SSD is successful up to sparsity $\rho \approx 0.072$ for Gaussian-type $\bm{h}^0$ and up to $\rho \approx 0.06$ for uniform-type $\bm{h}^0$. Comparing Fig.~\ref{fig:Cmtx10k} with Fig.~\ref{fig:CompOMP} we see that SSD performs almost equally good on the highly correlated matrix but OLS and OMP are very sensitive to correlations of the matrix.

\begin{figure}
  \centering
  \includegraphics[angle=270,width=1.0\columnwidth]{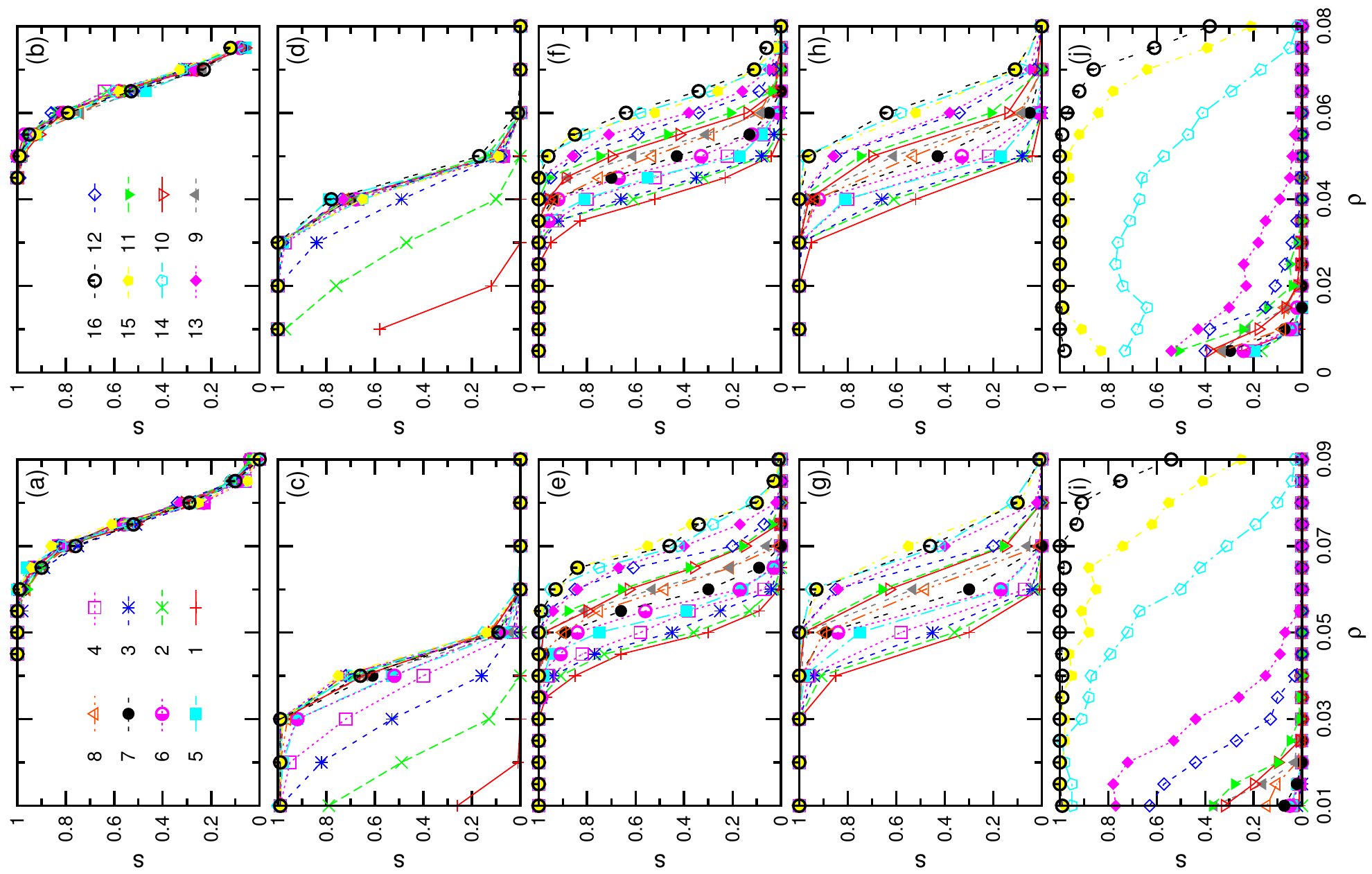}
  \caption{
    Comparing the algorithmic performances on $16$ correlated $200 \times 1000$ sampling matrices  $\bm{D}$ indexed from $1$ (pluses, most highly correlated) to $16$ (decorated circles, least correlated). Left column corresponds to Gaussian-type planted solutions $\bm{h}^0$, right column corresponds to uniform-type $\bm{h}^0$.  The simulation results are obtained by SSD [(a), (b)],  ML1 [(c), (d)], OLS [(e), (f)], OMP [(g), (h)] and AMP [(i), (j)], respectively. $\rho$ is the sparsity of $\bm{h}^0$; $s$ is the fraction of successfully reconstructed solutions among $100$ random input $\bm{h}^0$ samples.
  }
  \label{fig:Cor1k}
\end{figure}

To see how the algorithmic performance is affected by the condition number $Q$ of the sampling matrices, we generate a set of $16$ correlated $200 \times 1000$ matrices $\bm{D}$ according to the described protocol using different values of $R=200$, $208$, $220$, $240$, $260$, $280$, $300$, $350$, $400$, $450$, $500$, $700$, $1000$, $2000$, $3000$, $8000$. These matrices are labeled with index ranging from $1$ to $16$ in Fig.~\ref{fig:Cor1k} and their corresponding condition numbers are $Q=318.37$,  $101.68$,   $42.95$, $24.76$,   $17.72$,   $13.74$,   $10.93$, $8.96$,  $7.27$,    $6.45$,    $5.81$,    $4.47$, $4.00$,    $3.14$,    $2.97$ and  $2.73$, respectively. We find that SSD is very robust to correlations (Fig.~\ref{fig:Cor1k}a and ~\ref{fig:Cor1k}b) but the other algorithms are not: the performances of ML1, OLS and OMP, and AMP all severely deteriorate with the increasing of $Q$  (Fig.~\ref{fig:Cor1k}c - \ref{fig:Cor1k}j). 

We have also tested the algorithms on correlated matrices generated by some other more complicated protocols. These additional simulation results (not shown here) further confirm the high degree of insensitivity of SSD. This rather peculiar property should be very desirable in practical applications because it greatly relaxes the requirements on the sampling matrix.

\section{Conclusion and discussions}

In this paper we introduced the Shortest-Solution guided Decimation (SSD) algorithm for the compressed sensing problem and tested it on uncorrelated and correlated sampling matrices. The SSD algorithm is a geometry-inspired deterministic algorithm that does {\emph{not}} explicitly try to minimize a cost function. SSD outperforms OLS and OMP in recovering sparse planted solutions; and it is especially competitive in treating highly correlated or structured matrices, on which the tested other representative algorithms (ML1, OLS and OMP, and AMP) all fail.

Mathematical and algorithmic studies~\cite{Foucart-Rauhut-2013,Zhang-etal-2015} on the compressed sensing problem have focused overwhelmingly on uncorrelated random matrices satisfying the restricted isometric property~\cite{Candes-Tao-2006}. But the RIP incoherence condition can be severely violated in real-world practical problems (see \cite{Otsuki-etal-2017} for an example in physics on quantum Monte Carlo data analysis). Existing algorithms in the literature were not designed for tackling highly correlated sampling matrices, and this challenging issue has begun to be discussed only very recently (see, e.g., the work of \cite{Ma-Ping-2017,Rangan-etal-2016} on orthogonal AMP). The SSD algorithm is a significant step along this direction. The demonstrated high degree of tolerance to correlations indicates that SSD can serve as a versatile and robust tool for different types of compressed sampling problems and sparse representation/approximation problems. 

Rigorous theoretical understanding is largely absent on why the SSD algorithm is highly tolerant to structural correlations in the sampling matrix. We feel that (i) viewing the sparse recovery problem from the angle of eigen-subspace projection (\ref{eq:h0vsg}) and (ii) recursively adjusting this eigen-projection by modifying the matrix $\bm{D}$ and the signal vector $\bm{z}$ are crucial to make SSD insensitive to correlations. We offered some qualitative arguments in Section III, especially through (\ref{eq:h0vsg}) and (\ref{eq:gidecompose}), on why the dense guidance vector $\bm{g}$ is useful to sparse reconstruction.  More rigorous theoretical understanding on the SSD process needs to be pursued in the future. It looks quite promising to investigate the performance bound of SSD on ensembles of correlated matrices as a phase transition problem.

In this paper we only considered the ideal noise-free situation; but for practical applications it will be necessary to take into account possible uncertainty in the sampling matrix $\bm{D}$ and the unavoidable measurement noise in the signal vector $\bm{z}$. SSD is slower than OLS and OMP by a constant factor in terms of time complexity. To further accelerate the SSD process an easy adaptation is to fix a small fraction of the active indices instead of just one of them in each decimation step (see, for example, \cite{Wang-Li-2017}). We also need to extend the SSD algorithm to complex-valued compressed sensing problems.



\appendix

\section{Accelerated dual ascent process}
\label{sec:epsilon}

Let us denote by $\bm{\gamma}^{(t)}$ the gap vector after the $t$-th iteration step of the dual ascent process (\ref{eq:dam}), that is
\begin{equation}
  \bm{\gamma}^{(t)} \equiv \bm{z} - \bm{D} \bm{g}^{(t)} \; .
\end{equation}
Then after the $(t+1)$-th iteration step, we have
\begin{subequations}
  \begin{align}
    \bm{\beta}^{(t+1)} & = \bm{\beta}^{(t)} + \varepsilon^{(t)} \bm{\gamma}^{(t)}
    \; , \\
    \bm{\gamma}^{(t+1)} & =\bm{\gamma}^{t} - \varepsilon^{(t)} \bm{D} \bm{D}^T 
    \bm{\gamma}^{(t)} = \bm{\gamma}^{(t)} - \varepsilon^{(t)} \bm{\eta}^{(t)} \; ,
  \end{align}
\end{subequations}
where the auxiliary column vector $\bm{\eta}^{(t)}$ is defined as
\begin{equation}
  \bm{\eta}^{(t)} \equiv  \bm{D} (\bm{D}^T \bm{\gamma}^{(t)})\; .
\end{equation}
Notice that the Euclidean length ($\ell_2$-norm) of $\bm{\gamma}^{(t+1)}$ will be minimized by setting
\begin{equation}
  \varepsilon^{(t)} = \frac{\langle \bm{\gamma}^{(t)}, \bm{\eta}^{(t)} \rangle}{\langle \bm{\eta}^{(t)}, \bm{\eta}^{(t)}\rangle} \; .
\end{equation}
This optimal value of $\varepsilon^{(t)}$ is used in the dual ascent process.

\section{MATLAB code}
\label{sec:appendix}
 
As a simple demo we include here a MATLAB code which realizes the SSD algorithm in the most direct way. This code involves computing the pseudo-inverse of the matrix $\bm{D}$, so it is not optimal in terms of time complexity.  The more efficient implementation based on convex optimization (\ref{eq:convexmin}) is documented at {\tt power.itp.ac.cn/\~{}zhouhj/codes.html}.

\begin{lstlisting}[language=Matlab, ,frame=single,numbers=none,basicstyle=\footnotesize]
function [ h ] = ssd( D,z,epsilon )
    [m, n]=size(D);
    D0=D; z0=z;
    g=pinv(D)*z;
    support=[];
    remain=1:n;
    while norm(z)>epsilon
        [~,l]=max(abs(g));
        support=[support,remain(l)];
        a=D(:,l);
        a_unit=a/(a'*a);
        z=z-(z'*a)*a_unit ;
        D=D-repmat(a'*D,m,1).*repmat(a_unit,1,size(D,2));
        D(:,l)=[];
        remain(l)=[];
        g=pinv(D)*z;
    end
    h=zeros(n,1);
    h(support)=pinv(D0(:,support))*z0;
end

\end{lstlisting}

\section*{Acknowledgment}

H.J.Z.~thanks Jing He, Fengyao Hou and Hongbo Jin for help in computer simulations; P.Z.~acknowledges helpful discussions with Dong Liu, Xiangming Meng and Chuang Wang; M.S.~acknowledges the hospitality of ITP-CAS during his stay as an intern student.  H.J.Z.~and P.Z.~acknowledge the hospitality of Kavli Institute for Theoretical Sciences (KITS-UCAS) during the workshop ``Machine Learning and Many-Body Physics" (June 28--July 7, 2017).  This research was supported by the National Natural Science Foundation of China (grant numbers 11421063 and 11647601) and the Chinese Academy of Sciences (grant number QYZDJ-SSW-SYS018). The numerical computations were carried out at the HPC Cluster of ITP-CAS and the Tianhe-2 platform of the National Supercomputer Center in Guangzhou.



\end{document}